\documentclass[final,leqno,10pt,a4paper]{article}
\usepackage[a4paper, top=1in, bottom=1.25in, left=0.7in, right=0.7in]{geometry}
\usepackage[numbers,sort&compress]{natbib}
\usepackage{amsfonts}
\usepackage{amsmath}
\usepackage{amssymb}
  \usepackage{esint}
\usepackage[amssymb]{SIunits}
\usepackage[utf8]{inputenc}

\newcommand{\tens}{\mathbf}%
\newcommand{\vect}{\pmb}%

\DeclareMathOperator{\divv}{div}
\DeclareMathOperator{\tr}{tr}
\DeclareMathOperator{\Laplace}{\Delta}

\newcommand{\ud}{\mathrm{d}}

\newcommand{\defeq}{\stackrel{\mathrm{def}}{=}}

\newcommand{\Rey}{\mathrm{Re}}

\newcommand{\WOd}[3]{\vect{W}^{#1,#2}_{#3}(\Omega)}

\usepackage{tikz}
\usepackage{pgfplots}
\pgfplotsset{compat=1.13}

\usepackage{subfig}		
\usepackage[heightadjust=all,valign=b]{floatrow}
\usepackage{fr-subfig}
\usepackage{caption}
\newcommand{\Pcrit}{P_{\mathrm{crit}}}
\newcommand{\hatPcrit}{\hat{P}_{\mathrm{crit}}}
\newcommand{\onlyME}[1]{} %

\newcommand{\footME}[1]{} %

\newcommand{\FIXME}[1]{ \color{brown}[FIX: #1]\color{black}}

\newcommand{\Section}[1]{Section~\ref{#1}}
\newcommand{\Figure}[1]{Figure~\ref{#1}}

\usepgfplotslibrary{external} 
\newcommand{\monocolor}[2]{#1}
\monocolor{
	\pgfplotsset{
		axis background/.style={fill=black!5},
		unbounded coords=discard,
		every axis/.append style={
	 		width={0.48\textwidth}, 
			height={0.27\textheight},
			font=\footnotesize,
			mark size=1.5pt,
	 		cycle list name=mark list, %see p.195 of pgfplots.pdf
		},
		every axis legend/.append style={
			legend columns=1,
			font=\footnotesize,
	 		legend pos=north west,
		},
	}
}{%% color:
	\pgfplotsset{
		axis background/.style={fill=green!20},
		unbounded coords=discard,
		every axis/.append style={
	 		width={0.48\textwidth}, 
			height={0.27\textheight},
			font=\footnotesize,
			mark size=1.5pt,
	 		cycle list name=color, %see p.195 of pgfplots.pdf
		},
		every axis legend/.append style={
			legend columns=1,
			font=\footnotesize,
	 		legend pos=north west,
		},
	}
}
\newcommand{\dataSIRS}[2]{results/p01-#1-#2-90.txt}
\newcommand{\figureSIRSabcd}{
	\begin{figure}[t]
	\centering
 	\subfloat[flux $Q_{\mathrm{num}}$, for $\vect{b}=P_i\vect{n}$]{%
	\label{fig:steady-traction-Qs}
 		\centering
 		\begin{tikzpicture}
 		\begin{axis}[
 			xlabel=$P_1-P_2$, 	
			ylabel=$Q_{\mathrm{num}}$,
 			xmin=-1.0,	xmax=3.0,
 		]
 		\addplot+ table[x index=0,y index=1, smooth, solid, mark=.] {\dataSIRS{1}{1}};
 		\addplot+ table[x index=0,y index=1, smooth, solid, mark=.] {\dataSIRS{2}{1}};
 		\addplot+ table[x index=0,y index=1, smooth, solid, mark=.] {\dataSIRS{3}{1}};
 		\end{axis}
 		\end{tikzpicture}	
 	}%
	\hspace{0.04\textwidth}%
	\subfloat[non-radiality $i_{\mathrm{nr}}$, for $\vect{b}=P_i\vect{n}$]{%
	\label{fig:steady-traction-rads}
		\centering
		\begin{tikzpicture}
		\begin{axis}[
			xlabel=$P_1-P_2$,  
			ylabel=$i_{\mathrm{nr}}$,
			xmin=-1.0,	xmax=3.0,
			ymin=-0.05,
		]
		\addplot+ table[x index=0,y index=2, smooth, solid, mark=.] {\dataSIRS{1}{1}};
		\addlegendentry{ $Q_{\mathrm{init}}=0$ };
		\addplot+ table[x index=0,y index=2, smooth, solid, mark=.] {\dataSIRS{2}{1}};
		\addlegendentry{ $Q_{\mathrm{init}}=3$ };
		\addplot+ table[x index=0,y index=2, smooth, solid, mark=.] {\dataSIRS{3}{1}};
		\addlegendentry{ continuation };
		\end{axis}
		\end{tikzpicture}	
	}%
	\\
 	\subfloat[flux $Q_{\mathrm{num}}$, for $\hat{\vect{b}}=P_i\vect{n}$]{%
	\label{fig:steady-bhat-Qs}
 		\centering
 		\begin{tikzpicture}
 		\begin{axis}[
 			xlabel=$P_1-P_2$, 
 			ylabel=$Q_{\mathrm{num}}$,
 			xmin=-2.0,	xmax=2.0,
 		]
 		\addplot+ table[x index=0,y index=1, smooth, solid, mark=.] {\dataSIRS{1}{0}};
 		\addplot+ table[x index=0,y index=1, smooth, solid, mark=.] {\dataSIRS{2}{0}};
 		\addplot+ table[x index=0,y index=1, smooth, solid, mark=.] {\dataSIRS{3}{0}};
 		\end{axis}
 		\end{tikzpicture}	
 	}%
	\hspace{0.04\textwidth}%
	\subfloat[non-radiality $i_{\mathrm{nr}}$, for $\hat{\vect{b}}=P_i\vect{n}$]{%
	\label{fig:steady-bhat-rads}
		\centering
		\begin{tikzpicture}
		\begin{axis}[
			xlabel=$P_1-P_2$, 
			ylabel=$i_{\mathrm{nr}}$,
			xmin=-2.0,	xmax=2.0,
			ymin=-0.05,
			legend pos=south east,
		]
		\addplot+ table[x index=0,y index=2, smooth, solid, mark=.] {\dataSIRS{1}{0}};
		\addlegendentry{ $Q_{\mathrm{init}}=0$ };
		\addplot+ table[x index=0,y index=2, smooth, solid, mark=.] {\dataSIRS{2}{0}};
		\addlegendentry{ $Q_{\mathrm{init}}=3$ };
		\addplot+ table[x index=0,y index=2, smooth, solid, mark=.] {\dataSIRS{3}{0}};
		\addlegendentry{ continuation };
		\end{axis}
		\end{tikzpicture}	
	}%
	\caption{Numerical steady solutions in the isotropic radial setting: 
		the resulting flux $Q_{\mathrm{num}}$ (a,c) and non-radiality $i_{\mathrm{nr}}$ (b,d) 
		for given $P_i$--drops of the prescribed {\em traction} (a,b) or {\em full-gradient-traction} (c,d).
		The radial numerical solutions coincide with the analytical solutions.}
	\label{fig:steady-traction-bhat}
	\end{figure}
}

\newcommand{\pngSIRSdonot}[1]{01-nonrad-hand-#1.png}
\newcommand{\figureSIRSdonot}{
\begin{figure}[t]
	\centering
	\subfloat[velocity, $\vect{u}$]{%
		\includegraphics[width=0.48\textwidth,trim=20 0 40 0,clip=true]{\pngSIRSdonot{u}}
	}\hspace{0.01\textwidth}%
	\subfloat[pressure, $p$]{%
		\includegraphics[width=0.48\textwidth,trim=20 0 40 0,clip=true]{\pngSIRSdonot{p}}
	}
	\caption{Non-radial steady numerical solution (perfect-slip and {\em do-nothing} boundary conditions).}
	\label{fig:bhat-nonrad}
\end{figure}
}
\newcommand{\pngSJHdonot}[1]{03-hand-#1.pdf}
\newcommand{\figureSJHdonot}{
\begin{figure}[t]
	\centering
	\subfloat[velocity, $\vect{u}$]{%
		\includegraphics[width=0.48\textwidth,trim=20 0 40 0,clip=true]{\pngSJHdonot{u}}
	}\hspace{0.01\textwidth}%
	\subfloat[pressure, $p$]{%
		\includegraphics[width=0.48\textwidth,trim=20 0 40 0,clip=true]{\pngSJHdonot{p}}
	}
	\caption{Non-trivial steady numerical solution (no-slip and {\em do-nothing} boundary conditions).}
	\label{fig:bhat-jh}
\end{figure}
}

\newcommand{\dataUnJHS}[2]{p04/p04-ns0-Rey20.0-Pdrop#1.-Qinit#2.-m20.txt.plot}
\newcommand{\dataSJHS}[2]{results/p03-#1-#2-20.txt}
\newcommand{\figureJHab}{
\begin{figure}[t]
\centering
	\subfloat[steady solutions]{
	\label{fig:steady-JH}
		\centering
		\begin{tikzpicture}
		\begin{axis}[
			width={0.45\textwidth}, height={0.8\textwidth},
			xlabel=$P_1-P_2$,
			ylabel=$Q_{\mathrm{num}}$,
			xmin=-5.0, xmax=35.0,
			ymin=-2.0, ymax=5.0,
			legend pos=south east,
		]
		\addplot+ table[x index=0, y index=1, smooth, solid, mark=.] {\dataSJHS{1}{1}};
		\addlegendentry{ $\vect{b}=P_i\vect{n}$, $Q_{\mathrm{init}}=0$ };
		\addplot+ table[x index=0, y index=1, smooth, solid, mark=.] {\dataSJHS{2}{1}};
		\addlegendentry{ $\vect{b}=P_i\vect{n}$, $Q_{\mathrm{init}}=3$ };
		\addplot+ table[x index=0, y index=1, smooth, solid, mark=.] {\dataSJHS{1}{0}};
		\addlegendentry{ $\hat{\vect{b}}=P_i\vect{n}$, $Q_{\mathrm{init}}=0$ };
		\addplot+ table[x index=0, y index=1, smooth, solid, mark=.] {\dataSJHS{2}{0}};
		\addlegendentry{ $\hat{\vect{b}}=P_i\vect{n}$, $Q_{\mathrm{init}}=3$ };
		\end{axis}
		\end{tikzpicture}
	}%
	\hspace{0.06\textwidth}%
	\subfloat[unsteady solutions]{
	\label{fig:UnsteadyJH}
		\centering
		\begin{tikzpicture}
		\vspace{1em}%
		\begin{axis}[
			width={0.45\textwidth}, height={0.8\textwidth},
			xlabel={time $t$}, 
			ylabel=$Q_{\mathrm{num}}$,
			unbounded coords=discard,
			xmin=0.0, xmax=9,
			ymin=-2.5, ymax=8,
			legend pos=south east,
			cycle list name=linestyles
		]
		\addplot+ table[x index=0,y index=1, smooth] {results/\dataUnJHS{0}{-1}};
			\addlegendentry{ $P_1-P_2=0.$ };
		\addplot+ table[x index=0,y index=1, smooth] {results/\dataUnJHS{15}{-1}};
			\addlegendentry{ $P_1-P_2=15.$ };
		\addplot+ table[x index=0,y index=1, smooth] {results/\dataUnJHS{30}{-1}};
			\addlegendentry{ $P_1-P_2=30.$ };

		\pgfplotsset{cycle list shift=-3}
		\addplot+ table[x index=0,y index=1, smooth] {results/\dataUnJHS{0}{0}};
		\addplot+ table[x index=0,y index=1, smooth] {results/\dataUnJHS{15}{0}};
		\addplot+ table[x index=0,y index=1, smooth] {results/\dataUnJHS{30}{0}};
		\pgfplotsset{cycle list shift=-6}
		\addplot+ table[x index=0,y index=1, smooth] {results/\dataUnJHS{0}{1}};
		\addplot+ table[x index=0,y index=1, smooth] {results/\dataUnJHS{15}{1}};
		\addplot+ table[x index=0,y index=1, smooth] {results/\dataUnJHS{30}{1}};
		\pgfplotsset{cycle list shift=-9}
		\addplot+ table[x index=0,y index=1, smooth] {results/\dataUnJHS{0}{2}};
		\addplot+ table[x index=0,y index=1, smooth] {results/\dataUnJHS{15}{2}};
		\addplot+ table[x index=0,y index=1, smooth] {results/\dataUnJHS{30}{2}};
		\pgfplotsset{cycle list shift=-12}
		\addplot+ table[x index=0,y index=1, smooth] {results/\dataUnJHS{0}{3}};
		\addplot+ table[x index=0,y index=1, smooth] {results/\dataUnJHS{15}{3}};
		\addplot+ table[x index=0,y index=1, smooth] {results/\dataUnJHS{30}{3}};
		\pgfplotsset{cycle list shift=-15}
		\addplot+ table[x index=0,y index=1, smooth] {results/\dataUnJHS{0}{4}};
		\addplot+ table[x index=0,y index=1, smooth] {results/\dataUnJHS{15}{4}};
		\addplot+ table[x index=0,y index=1, smooth] {results/\dataUnJHS{30}{4}};
		\end{axis}
		\end{tikzpicture}
	}%
	\caption{Numerical solutions for Jeffery--Hamel flow: 
	(a) the resulting flux $Q_{\mathrm{num}}$ of the steady flow,
	for various {\em traction} or {\em full-gradient-traction} $P_i$--drops 
	and two initial guesses; and
	(b) the resulting flux $Q_{\mathrm{num}}(t)$ of the unsteady flow,
	for three {\em full-gradient-traction} $P_i$--drops
	and five initial states.}
	\end{figure}
}

\newcommand{\pdfHemo}[1]{05-02-#1.png}
\newcommand{\figureHemo}{
\begin{figure}[t]
	\centering
	\subfloat[obtained from trivial initial guess]{%
	\label{fig:hemo-a}%
		\includegraphics[width=0.47\textwidth,trim=15 0 5 0,clip=true]{\pdfHemo{0}}
	}\hspace{0.03\textwidth}%
	\subfloat[obtained from non-trivial initial guess]{%
	\label{fig:hemo-b}%
		\includegraphics[width=0.47\textwidth,trim=15 0 5 0,clip=true]{\pdfHemo{-7}}
	}
	\caption{Two numerical solutions ({\em do-nothing} conditions on two outlets).}
	\label{fig:hemo}
\end{figure}
}

\begin{document}
\title{%
On multiple solutions to the steady flow of incompressible fluids
subject to do-nothing or constant traction boundary conditions on artificial boundaries%
\thanks{%
Submitted to Journal of Mathematical Fluid Mechanics. 
The final publication is available at Springer via 
\href{https://doi.org/10.1007/s11012-017-0731-0}{https://doi.org/10.1007/s11012-017-0731-0}.%
}%
}%

\author{M. Lanzend{\"o}rfer%
\thanks{%
Charles University, Faculty of Science, %
Albertov 6, Prague 2, Czech Republic,
{\tt martin.lanzendorfer@natur.cuni.cz}}
\and
J. Hron%
\thanks{Charles University, Faculty of Mathematics and Physics, %
Sokolovsk{\'a} 49/83, Prague 8, Czech Republic,
{\tt hron@karlin.mff.cuni.cz}}
}

\date{April 9, 2019}

\maketitle
\begin{abstract}
The boundary conditions prescribing the constant traction or the so-called {\em do-nothing} conditions
are frequently taken on artificial boundaries in the numerical simulations
of steady flow of incompressible fluids, despite the fact that they
do not facilitate a~well-posed problem, not allowing to establish the standard energy estimates. 
In a~pursuit to understand better the possible consequences of using these
conditions, we present a~particular set of examples of flow problems, where we find 
none or two analytical or numerical solutions.
Namely, we consider problems where the flow connects two such artificial boundaries.
In the simple case of the isotropic radial flows %
where both steady and unsteady analytical solutions are derived easily, %
we demonstrate that while for some (large) boundary data all unsteady solutions blow up in finite time,
for other data (including small or trivial) %
the unsteady flows either converge asymptotically to one of two steady solutions,
or blow up in finite time, depending on the initial state. 
We then document the very same behavior of the numerical solutions for planar flow in a~diverging channel.
Finally, we provide an~illustrative example of two steady numerical solutions to the flow 
in a~three-dimensional bifurcating tube, where the inflow velocity is prescribed at the inlet, 
while the two outlets are treated by the {\em do-nothing} boundary condition.

\end{abstract}

\section{Introduction}\label{sec:intro}

The most of studies in fluid dynamics or computational fluid dynamics 
deal with flows in domains smaller than the whole space $\mathbb{R}^2$ or $\mathbb{R}^3$.
The governing equations of the flow then consist of a~system of partial differential equations %
inside the domain $\Omega$ and additional conditions given on the boundary $\partial\Omega$,
only together can they define a~well-posed problem.
On the parts of the boundary that are related to a~natural material interface, 
the physics of the interface is employed 
to derive the boundary conditions, similarly as when deriving
the partial differential equations in the bulk based on the rheology of the fluid. 
The no-slip assumption considered on the interface of the fluid with a~solid wall 
and the corresponding Dirichlet boundary conditions
can serve as an~elementary example, one among others.
In many practical problems however, 
some parts of the boundary are not related to any such interface
but represent a~truncation of the flow extending beyond the considered domain;
they are artificial boundaries.
The boundary conditions on artificial boundaries cannot be derived merely from the
underlying physics, but are justified by a~combination of arguments related to 
the physics, the modelling goals, the well-posedness analysis and the numerical analysis.

This paper is concerned with two types of boundary conditions that are frequently used  
in numerical simulations of flows of incompressible fluids, in particular Navier--Stokes fluid, 
despite the fact that they do not facilitate the well-posedness of the problem.
Namely, as introduced in a~greater detail below, the boundary conditions prescribing a~constant {\em traction},
or the ones that we later call the {\em full-gradient-traction} conditions and cover the so-called {\em do-nothing}
conditions, are considered.
It is well known that the steady Navier--Stokes equations subject to these %
boundary conditions do not allow for standard energy estimates.
Consequently, 
it has been impossible so far %
to establish the existence theory for the steady problem except for small data
(and similarly, 
to establish the existence of unsteady solution except for small data or small time interval).
While the non-uniqueness of steady solutions seems to be expected intuitively, 
no concrete example of multiple solutions has been given in the literature so far,
to the best of our knowledge.
Indeed, as Galdi comments in \cite{Galdi2008-inHemodynamicalFlows}, p.~180, %
(after proving that for small data there is a~unique {\em small} solution)
\begin{quotation}
	``\dots the question of whether a~given solution is unique in the class of all possible
	weak solutions corresponding to the same data 
	\dots 
	is, to date, open, in the case of the {\em do-nothing} conditions.''
\end{quotation}
and similarly does Rannacher in \cite{Rannacher2008-inHemodynamicalFlows}, p.~294,
\begin{quotation}
	``Unfortunately, in the case of open outlets such an {\em a~priori} bound is not known. 
	This is reflected by the fact that not even the global uniqueness of the zero-solution
	has been proven yet \ldots
	However, this possible non-uniqueness could not be confirmed by numerical experiments.''
\end{quotation}
\onlyME{ MOVE THIS:
The standard remedy is to modify the boundary conditions by an additional term
related to the kinetic energy brought into the domain by eventual inflow.
Such modification, however, brings certain disadvantages.
Most importantly, the modified conditions are not satisfied by simple flows such as Poiseuille or Couette flow.
}
The very last sentence of the above quotation points out an important aspect:
that, in fact, the addressed boundary conditions are used
in numerical simulations by many researchers on a~regular basis, 
since in many problem geometries and flow regimes they are experienced to deliver a~unique solution.

This paper is motivated by the above stated 
incongruity between the hitherto achievements in the theoretical analysis
of Navier--Stokes equations, and the well-established techniques and needs in practical numerical simulations.
Our aim is to contribute to a~better understanding of the corresponding solutions,
or even to provoke further study of the theoretical aspects of 
Navier--Stokes equations in this respect. %
The scope of the paper is to present a~set of examples of multiple solutions 
to the steady flow subject to the above addressed boundary conditions.
Aside from the mere confirmation of the non-uniqueness of solutions 
we intend to demonstrate a~particular simple mechanism behind, 
and to pursue the behavior of the steady solutions, of the corresponding unsteady flows
and of their numerical approximation.
We observe multiple solutions for small boundary data,
including the case of trivial data where both the~trivial and non-trivial solutions can be found.
On the other hand, for some instances of large boundary data, the considerations in a~simplified setting 
and our numerical simulations indicate the possibility that no steady solution would exist,
this fact being related to the very same mechanism.

\subsection{Weak solutions, traction, full-gradient-traction, and do-nothing}
The steady Navier--Stokes equations can be written as
\begin{equation}\label{eq:NS}
\left.
\begin{array}{rcl}
		\divv(\vect{u}\otimes\vect{u})
	-	\divv\tens{T}
	&=&	\vect{f}
	\\	\divv\vect{u} &=& 0
\end{array}
\quad\right\}
\quad	\textrm{ in $\Omega$, where}	\quad
\begin{array}{rcl}
	\tens{T} &=& -p\tens{I} + \tens{S} %
,\\	\tens{S} %
	&=& \nu\,(\nabla\vect{u}+(\nabla\vect{u})^T)
.\end{array}
\end{equation}
Here $p$, $\vect{u}$ denote the kinematic pressure and the velocity, 
$\tens{T}$ and $\tens{S}$ stand for Cauchy stress tensor and its viscous part,
the constant $\nu>0$ being the kinematic viscosity,
and $\vect{f}$ represents the density of outer forces 
(later on, we set $\vect{f}\equiv\vect{0}$ for simplicity).
The domain $\Omega\subset\mathbb{R}^2$ or $\Omega\subset\mathbb{R}^3$ is considered bounded
and with Lipschitz boundary $\partial\Omega\in\mathcal{C}^{0,1}$.

The weak formulation can be formally derived by multiplying the system \eqref{eq:NS} by suitable 
test functions $\vect{w}$, $q$ and then integrating over~$\Omega$.
One arrives at 
\begin{align*}
	(\, \divv(\vect{u}\otimes\vect{u}) , \vect{w} \,)_\Omega
-	(\, \divv\tens{T} , \vect{w} \,)_\Omega
&=	(\, \vect{f} , \vect{w} \,)_\Omega
,\\	(\, \divv\vect{u} , q \,)_\Omega 
&=	0
,\end{align*}
to hold for all $\vect{w}$, $q$.
The suitable function spaces for $\vect{u}$, $\vect{w}$ and $p$, $q$ are to be specified 
depending on the boundary conditions considered.
By the brackets we shortly denote the integrals of the scalar product 
$(\vect{f},\vect{g})_\Omega=\int_\Omega \vect{f}\cdot\vect{g}\,\mathrm{d}\vect{x}$,
and analogously for scalars or tensors and for the integration on $\partial\Omega$.
The divergence theorem allows us to write
$$	(\, \divv\tens{T} , \vect{w} \,)_\Omega
=	(\, \tens{T}\vect{n} , \vect{w} \,)_{\partial\Omega}
-	(\, \tens{T} , \nabla\vect{w} \,)_\Omega
,$$
where $\vect{n}$ denotes the unit outer normal vector.
From this it appears that the weak formulation offers two quantities to be naturally considered on the boundary:
the velocity $\vect{u}$ (to be treated by an~{\em essential}, Dirichlet, boundary condition)
and the surface traction $-\tens{T}\vect{n}$ (to be imposed by a~{\em natural}, Neumann, boundary condition).
Imposing, for instance, the no-slip conditions on a~part of the boundary 
\begin{equation}\label{eq:noslip}
	\vect{u} = \vect{0}
\qquad	\textrm{on $\Gamma_0\subset\partial\Omega$}
,\end{equation}
and prescribing the traction $\vect{b}$ on the remaining part (the artificial boundary)
\begin{equation}\label{eq:b}
\begin{array}{rcl}
	-\tens{T}\vect{n} 
	= p\vect{n} - \tens{S}\vect{n}
	=&&
\\	p\vect{n} - \nu\,(\nabla\vect{u}+(\nabla\vect{u})^T)\vect{n} 
	&=& 	\vect{b}
\qquad	\textrm{on $\Gamma_{\vect{b}}\subset\partial\Omega$}
,\end{array}
\end{equation}
the weak solution can be defined for example as follows:
{\em
(For $\Omega$, $\partial\Omega=\Gamma_0\cup\Gamma_{\vect{b}}$,
$\nu>0$, and 
$\vect{f}\in\vect{L}^2(\Omega)$, $\vect{b}\in\vect{L}^2(\Gamma_{\vect{b}})$ given)
find $\vect{u}\in\WOd{1}{2}{\Gamma_0}$ and $p\in L^2(\Omega)$,
such that %
for all $\vect{w}\in\WOd{1}{2}{\Gamma_0}$ and $q\in L^2(\Omega)$
\begin{equation}\label{eq:NSb}
	(\, \divv(\vect{u}\otimes\vect{u}) , \vect{w} \,)_\Omega
+	(\, \nu\,( \nabla\vect{u}+(\nabla\vect{u})^T) , \nabla\vect{w} \,)_\Omega
-	(\, p , \divv\vect{w} \,)_\Omega
-	(\, q , \divv\vect{u} \,)_\Omega 
=	(\, \vect{f} , \vect{w} \,)_\Omega
- 	(\, \vect{b} , \vect{w} \,)_{\Gamma_{\vect{b}}}
.\end{equation}%
}%
Here $\WOd{1}{2}{\Gamma_0} = \{ \vect{v}\in\WOd{1}{2}{} \,,\; \tr\vect{v}=\vect{0} \textrm{ on }\Gamma_0 \}$
is the standard Sobolev space of functions with $L^2$-integrable derivatives in $\Omega$ and the values vanishing on~$\Gamma_0$.
All the data are considered $L^2$-integrable here for the sake of brevity.
It is worth noting that the above procedure can be used for fluids exhibiting variable viscosity as well; 
the boundary condition \eqref{eq:b} would retain the physical meaning of prescribing the traction on $\Gamma_{\vect{b}}$.

Since the viscosity is constant,
the incompressibility constraint $\divv\vect{u}=0$ allow us to write 
$$	\divv(\nu\,(\nabla\vect{u}+(\nabla\vect{u})^T)\,)
=	\nu\divv(\nabla\vect{u})  + \nu\nabla(\divv\vect{u})
=	\nu\divv(\nabla\vect{u})
=	\nu\Laplace\vect{u}
.$$
The classical form of the momentum equation \eqref{eq:NS}$_1$ is thus more usually written as 
$$
		\divv(\vect{u}\otimes\vect{u})
	-\nu	\Laplace\vect{u}
	+	\nabla p
	=	\vect{f}
\qquad	\textrm{in $\Omega$}
.$$
Correspondingly, one can write \eqref{eq:NS} equivalently with $\tens{T}$ 
replaced formally by 
$$	\hat{\tens{T}}=-p\tens{I}+\hat{\tens{S}}
,\qquad \hat{\tens{S}}=\nu\,\nabla\vect{u}
,$$
and follow the above procedure, i.e.\ apply the divergence theorem on $\divv\hat{\tens{T}}$,
the quantity naturally emerging on the boundary being now $-\hat{\tens{T}}\vect{n}$, instead of the traction.
Thus, one is led to replace the boundary condition \eqref{eq:b}
by its analogy
\begin{equation}\label{eq:bhat}
\begin{array}{rcl}
	-\hat{\tens{T}}\vect{n} 
	= p\vect{n} - \hat{\tens{S}}\vect{n}
	=&&
\\	p\vect{n} - \nu\,(\nabla\vect{u})\vect{n} 
	&=& 	\hat{\vect{b}}
\qquad	\textrm{on $\Gamma_{\hat{\vect{b}}}\subset\partial\Omega$}
.\end{array}
\end{equation}
In the special case that $\hat{\vect{b}}\equiv\vect{0}$ 
this condition is well-known as the {\em do-nothing} boundary condition, 
see~\cite{HeywoodRannacherTurek1996} for a~detailed discussion.
In the case that $\hat{\vect{b}}\neq\vect{0}$ in general%
\footnote{
	By the course of history, the label {\em do-nothing} usually refers to the condition 
	\eqref{eq:bhat} with $\hat{\vect{b}}=\vect{0}$,
	but not to the zero traction $\vect{b}=\vect{0}$ in \eqref{eq:b}.
	The term is not related to the underlying physics in any way.
	The label originates from the mere fact 
	that the boundary term $(\hat{\vect{b}},\vect{w})_{\partial\Omega}$ 
	is absent in the weak formulation \eqref{eq:NSbhat}.
	Therefore, we do not consider it convenient to use the term {\em do-nothing}
	for the condition \eqref{eq:bhat} with general $\hat{\vect{b}}$.
}%
, we take the liberty to call \eqref{eq:bhat}
the {\em full-gradient-traction} condition, for the purpose of this paper.
The weak solution corresponding to \eqref{eq:noslip} and \eqref{eq:bhat}
is then defined as follows:
{\em
(For $\Omega$, $\partial\Omega=\Gamma_0\cup{\Gamma}_{\hat{\vect{b}}}$,
$\nu>0$, and 
$\vect{f}\in\vect{L}^2(\Omega)$, $\hat{\vect{b}}\in\vect{L}^2(\Gamma_{\hat{\vect{b}}})$ given)
find $\vect{u}\in\WOd{1}{2}{\Gamma_0}$ and $p\in L^2(\Omega)$,
such that 
for all $\vect{w}\in\WOd{1}{2}{\Gamma_0}$ and $q\in L^2(\Omega)$
\begin{equation}\label{eq:NSbhat}
	(\, \divv(\vect{u}\otimes\vect{u}) , \vect{w} \,)_\Omega
+	(\, \nu\nabla\vect{u} , \nabla\vect{w} \,)_\Omega
-	(\, p , \divv\vect{w} \,)_\Omega
-	(\, q , \divv\vect{u} \,)_\Omega 
=	(\, \vect{f} , \vect{w} \,)_\Omega
- 	(\, \hat{\vect{b}} , \vect{w} \,)_{\Gamma_{\hat{\vect{b}}}}
.\end{equation}%
}%

\subsection{Lack of energy estimates, possible remedy and simple flows}
The energy estimates for the weak solutions to the steady Navier--Stokes equations
are obtained by picking the velocity solution as the test function in the weak momentum equation.
By taking $\vect{w}=\vect{u}$ in \eqref{eq:NSb} or \eqref{eq:NSbhat}
and using that $\divv\vect{u}=0$ a.e.\ in~$\Omega$,
one arrives at
\begin{align*}
	(\, \divv(\vect{u}\otimes\vect{u}) , \vect{u} \,)_\Omega
+	\left( \nu\, (\nabla\vect{u}+(\nabla\vect{u})^T), \frac{\nabla\vect{u}+(\nabla\vect{u})^T}{2} \right)_\Omega
&=	(\, \vect{f} , \vect{u} \,)_\Omega
- 	(\, \vect{b} , \vect{u} \,)_{\Gamma_{\vect{b}}}
,\\
\textrm{or}\qquad\qquad %
	(\, \divv(\vect{u}\otimes\vect{u}) , \vect{u} \,)_\Omega
+	(\, \nu\, \nabla\vect{u}, \nabla\vect{u} \,)_\Omega
&=	(\, \vect{f} , \vect{u} \,)_\Omega
- 	(\, \hat{\vect{b}} , \vect{u} \,)_{\Gamma_{\hat{\vect{b}}}}
,\end{align*}
respectively, allowing for the estimate 
\begin{equation}\label{eq:estimate}	
	(\, \divv(\vect{u}\otimes\vect{u}) , \vect{u} \,)_\Omega
+	\sigma \| \tilde{\tens{S}} \|_{2,\Omega}^2
\leq	
	\| \vect{f} \|_{2,\Omega}  \|\vect{u}\|_{2,\Omega}
+	\| \tilde{\vect{b}} \|_{2,\Gamma_{\tilde{\vect{b}}}} \|\vect{u}\|_{2,\Gamma_{\tilde{\vect{b}}}}
,\end{equation}
defining either 
$\tilde{\tens{S}} \defeq \tens{S}$, 		$\tilde{\vect{b}} \defeq \vect{b}$ and 		
$\Gamma_{\tilde{\vect{b}}}\defeq\Gamma_{\vect{b}}$ from \eqref{eq:b}
and $\sigma \defeq (2\nu)^{-1}$,
or defining 
$\tilde{\tens{S}} \defeq \hat{\tens{S}}$, 	$\tilde{\vect{b}} \defeq \hat{\vect{b}}$ and
$\Gamma_{\tilde{\vect{b}}}\defeq\Gamma_{\hat{\vect{b}}}$ from \eqref{eq:bhat}
and $\tilde{\nu} \defeq {\nu}^{-1}$.
Here $\|\cdot\|_{2,\Omega}$ denotes the standard $L^2$-norm in $\Omega$
and analogously%
\footnote{%
	We only take $L^2$-estimates of the boundary integrals, for simplicity.
}%
on $\partial\Omega$.
The cubic convective term integral is subject to the following identity, using again that $\divv\vect{u}=0$,
$$	(\, \divv(\vect{u}\otimes\vect{u}) , \vect{u} \,)_\Omega
=	\tfrac12 \,(\, \divv\vect{u} , |\vect{u}|^2 \,)_\Omega
+	\tfrac12 \,(\, \vect{u}\cdot\vect{n} , |\vect{u}|^2 \,)_{\partial\Omega}
=	\tfrac12 \,(\, \vect{u}\cdot\vect{n} , |\vect{u}|^2 \,)_{\Gamma_{\tilde{\vect{b}}}}
,$$
and is thus determined by the values of velocity on the boundary;
its negative represents the inflow rate of the kinetic energy coming through the artificial boundary.
In the case of Dirichlet boundary conditions imposed along the entire boundary
(or any slip conditions, where $\vect{u}\cdot\vect{n}=0$ on $\partial\Omega$),
the cubic term would vanish completely and one would obtain an a~priori bound for~$\tilde{\tens{S}}$,
concluding the energy estimate for $\vect{u}$ in $\WOd{1}{2}{}$ in terms of the data~$\vect{f}$
(by using Poincar{\'e}'s or Korn's inequality, with further requirements on~$\Gamma_0$).
In contrast, the kinetic energy term does not vanish if the artificial boundaries are considered,
since the normal component of the velocity is not under control.
In this way, for the boundary conditions \eqref{eq:b} or \eqref{eq:bhat}
with the given boundary data $\tilde{\vect{b}}\equiv\tilde{\vect{b}}(\vect{x})$,
one cannot derive the {\em a~priori} energy bounds.

This fact is reflected by the lack of theoretical results that would guarantee the existence 
and uniqueness of a~steady weak solution. 
The existence has been stated%
\footnote{%
	Here we only follow the main idea and leave the details on how the results are formulated to the reader.
	For instance, the referred results are not stated for general $\tilde{\vect{b}}=\tilde{\vect{b}}(\vect{x})$,
	but rather for the particular choice $\hat{\vect{b}}=P_i\vect{n}$.
} only for the case of small data in~\cite{HeywoodRannacherTurek1996},
while without such requirement it remains an~open problem in both two and three space dimensions, 
see also \cite{Rannacher2012,BraackMucha2014}.
A~detailed discussion can be found in \cite{Galdi2008-inHemodynamicalFlows}, 
where the uniqueness of a {\em small} steady weak solution (for small data) was shown. %
The uniqueness in the class of {\em all} weak solutions 
(even the uniqueness of the trivial solution for the trivial data) 
is considered an~open problem, as quoted in the first subsection.
Similarly, concerning the non-stationary Navier--Stokes problem involving the boundary conditions 
\eqref{eq:b} or \eqref{eq:bhat}, the available results only guarantee the existence and uniqueness
of weak solutions either for small data or for a~small time interval, 
see~\cite{Galdi2008-inHemodynamicalFlows,BotheKohnePruss-2013} and the references therein.
In numerical simulations of unsteady flow, this is experienced as a~severe instability which 
appears for larger Reynolds numbers whenever a~backflow through such boundary 
occurs (either through an~entire artificial boundary part,
or locally, e.g.\ due to vortices leaving the domain through the outlet)
and which may quickly destroy the numerical solution,
see also~\cite{BertoglioEtal-2018}.

A~number of techniques has been developed to avoid the above mentioned effects in numerical simulations,
in particular in the context of unsteady flow, where they are often referred to as ``stabilization'' methods.
We refer the reader to~\cite{BertoglioEtal-2018} for a~selection of various approaches with references. 
The idea behind one class of these techniques,
to illustrate the point briefly, 
is to modify the boundary conditions \eqref{eq:b} or \eqref{eq:bhat}
so that the kinetic energy of the eventual inflow (backflow) is compensated.
This requires to take the boundary data
$\tilde{\vect{b}}\equiv\tilde{\vect{b}}(\vect{x},\vect{u})$ 
such that,
keeping the notation same as above,
$$	
	\int_{\Gamma_{\tilde{\vect{b}}}}
		\tilde{\vect{b}}(\vect{u})\cdot\vect{u} 
	\,\ud S
\geq 	-\frac12 \int_{\Gamma_{\tilde{\vect{b}}}} 
		(\vect{u}\cdot\vect{n}) |\vect{u}|^2
	\,\ud S
	- \textrm{ lower order terms}
$$
holds for all admissible functions $\vect{u}$, say for all $\vect{u}\in\vect{L}^3(\Gamma_{\tilde{\vect{b}}})$.
In the case of outflow, the cubic term on the right hand side is negative and does not represent
any additional restriction on $\tilde{\vect{b}}$.
However, since the occurrence of an~inflow ($\vect{u}\cdot\vect{n}<0$) cannot be precluded,
a~suitable quadratic dependence of the  boundary data $\tilde{\vect{b}}$ on the velocity,
such as for instance
\begin{equation}\label{eq:bReys}
	\tilde{\vect{b}}(\vect{u}) = P_0\vect{n} - \frac12(\vect{u}\cdot\vect{n})\vect{u}
\qquad	\textrm{or}
\qquad	\tilde{\vect{b}}(\vect{u}) = \left(P_0 - \frac12 |\vect{u}|^2 \right)\vect{n}
\end{equation}
is required.
In~\cite{BruneauFabrie1996}, this approach allowed to show the existence of unsteady solutions
for arbitrarily large data, and has been utilized in a~number of theoretical works since then
(see, e.g., Sect.\ 5 in~\cite{LanzStebel2011a} for a~list of further references).
This includes the result in \cite{TNeustupa2016} showing 
the existence of a~steady weak solution with an~arbitrarily large prescribed inflow 
(imposed via the Dirichlet boundary condition $\vect{v}=\vect{g}$, with $\vect{g}$ given on some part of $\partial\Omega$)
subject to the outflow condition
\begin{equation}\label{eq:DDN}
	\hat{\vect{b}}(\vect{x},\vect{u}) = \vect{h}(\vect{x}) + \alpha\,(\vect{u}\cdot\vect{n})^- \vect{u}
\qquad	\textrm{for } 
	\alpha>\frac12
,\end{equation}
where $0\leq(\vect{u}\cdot\vect{n})^-$ denotes the negative part of $\vect{u}\cdot\vect{n}$,
see also the discussion and the numerical experiments presented in~\cite{BraackMucha2014}.
Analogous results can be shown for fluids with variable viscosity,
see for instance \cite{LanzStebel2011a},
where the steady flow of fluids with the pressure- and shear rate-dependent viscosity was considered.

\onlyME{ je jeste mnoho dalsich pristupu, see Dong, a citace v Benchmark paperu}
\onlyME{
An~alternative approach was introduced by~\cite{KracmarNeustupa1994,KracmarNeustupa2001},
where a~variational inequality based on Navier--Stokes system is considered, 
\FIXME{ supplementing; tak plati nebo ne?}  the boundary condition \eqref{eq:bhat} by the additional constraint
$$	\int_{\Gamma_{\hat{\vect{b}}}} |(\vect{u}\cdot\vect{n})^-|^a \,\ud S \leq c_0
,\qquad	\textrm{ for some $a$, $c_0>0$}
.$$
This again allows to derive the energy estimates,
eventually proving the existence of solutions
(which are not guaranteed to meet \eqref{eq:bhat}, however).
}

While the above mentioned modified boundary conditions allow to define problems that are 
mathematically well-posed, it seems that the conditions \eqref{eq:b} and \eqref{eq:bhat}
are those that are better established and more frequently used in 
the steady flow numerical simulations,
and even for unsteady flows as far as the backflow is not expected
or Reynolds number is small.
In fact, despite the lack of rigorous results, they both seem quite well-behaved 
in many problem settings.
Note that the {\em full-gradient-traction} condition \eqref{eq:bhat}
provides a~property that all the others mentioned above lack: 
The condition
\begin{equation*}%
	\hat{\vect{b}} = P_0 \vect{n}
,\qquad	\textrm{i.e.}
\qquad	(p-P_0)\vect{n} - \nu\,(\nabla\vect{u})\vect{n} = \vect{0}
,\end{equation*}
is satisfied by unidirectional simple flows such as Couette or Poiseuille flow,
on the inflow and outflow boundaries (defined so that the normal vector $\vect{n}$ is parallel to the flow).
Here $P_0\in\mathbb{R}$ is a~given constant and, in the resulting simple flow, 
$p=P_0$ on $\Gamma_{\hat{\vect{b}}}$.
As has been discussed in \cite{HeywoodRannacherTurek1996} already,
the boundary conditions such as \eqref{eq:bReys} 
result in the velocity solutions with the streamlines distorted near the artificial boundary
and the corresponding artefacts in the pressure field.
These effects are in many situations undesirable, %
or are even referred to by some authors as ``unphysical'', 
in particular in the case of flows in channels or pipes, or e.g.\ in blood vessels, 
where one naturally supposes that the flow continues undisturbed beyond the artificial boundary.
While the %
``directional'' variants such as \eqref{eq:DDN} avoid this disadvantage in the case of outflow,
it suffers the same disability in problems, 
where the inflow boundary needs to be treated in the same way
(for one such example, see the plane slider problem studied in~\cite{LanzMalekRaj-2018}).

\subsection{The structure of the paper}\label{sec:structure}
We present a~particular set of flow problems in bounded domains in two or three space dimensions, 
where one finds multiple steady solutions subject to the same boundary data,
the data of the problem being arbitrarily small or even trivial. 
The specific feature of the presented examples is that 
the constant {\em traction} \eqref{eq:b} or {\em full-gradient-traction} \eqref{eq:bhat} 
conditions are prescribed on two%
\footnote{%
	One can experiment with more artificial boundaries as well, see \Section{sec:other}.
} 
boundaries 
$\Gamma_i\subset\partial\Omega$ (one opposite to another),
i.e.
$$	\tilde{\vect{b}} = P_i \vect{n}
\qquad	\textrm{on } \Gamma_i
,\quad	i=1,2
,$$ 
where $\tilde{\vect{b}}=\vect{b}$ or  $\tilde{\vect{b}}=\hat{\vect{b}}$
and where $P_i\in\mathbb{R}$ are given constants on each $\Gamma_i$.
The particularly appealing case where the {\em do-nothing} condition $\hat{\vect{b}}=\vect{0}$
is imposed on all artificial boundaries is included.

In order to reveal the simple mechanism which promotes this multiplicity, we start 
in \Section{sec:radial} by studying the isotropic radial planar flow,
where the artificial boundaries are considered at the radii $0<R_1<R_2$.
Given the {$P_i$--drop}%
\footnote{%
In unidirectional simple flows, such as the flows in channels or pipes,
	the value $P_i$, both whether given as the normal {\em traction} or {\em full-gradient-traction},
	would correspond to the resulting pressure.
	It is then natural to call the difference $P_1-P_2$ the {\em pressure drop}
	and to speak about {\em pressure drop} problems.
	However, it is worth noting that in non-unidirectional flows, such as the radial flows,
	the data $P_i$ do {\em not} coincide with the resulting pressure.
	For the sake of brevity, let us denominate $P_1-P_2$ as {$P_i$--drop}, in this paper.
} $P_1-P_2$, 
the steady problem then reduces to the task of finding one constant,
the flow rate $Q\in\mathbb{R}$, which appears to be the solution of a~quadratic equation.
Apparentally, there is a~critical value $\Pcrit>0$, such that 
there are two such steady solutions for $P_1-P_2<\Pcrit$, 
which includes the case of trivial boundary data, %
while for $P_1-P_2>\Pcrit$ there is no (isotropic radial) steady solution.
Taking the advantage of this reduced setting, we continue in \Section{sec:stability}
by studying the unsteady problem (for stationary data). %
The isotropic radial solution is then given by a~single function of time, $Q(t)$,
which is found by solving the corresponding ordinary differential equation.
We find that the solution either converges asymptotically to one of the steady solutions
or it blows up in finite time.

\Section{sec:fem} is then devoted to the finite element approximation of the flow.
We start by examining the isotropic radial setting in \Section{sec:fem-radial},
defining the flow domain as one quadrant of a~concentric annulus, where 
the symmetry (or, say, perfect-slip) conditions are imposed on the straight boundaries.
We observe that the common simple scheme based on Newton's method can find 
both of the two steady solutions, depending on the given initial guess.
For large {$P_i$--drops}, a~numerical steady solution is found, but it is non-radial.
Examining the unsteady case, we confirm numerically the behavior found analytically,
including the blow-up of the unsteady solutions in finite time.

Doubt may arise, whether the solutions in the isotropic radial setting 
are related to the problems where the {\em do-nothing} conditions are used most: 
to the channel-like flows, where the supposed outflow is nearly a~unidirectional simple flow.
In \Section{sec:fem-JH}, we thus make the domain $\Omega$ more narrow and 
impose no-slip conditions on the walls, studying what resembles a~Jeffery--Hamel flow.
The numerical results are qualitatively same as in the isotropic radial setting, 
except that we are unable to find any steady solution for large {$P_i$--drops}.

Ultimately, in \Section{sec:bif} we study the three-dimensional flow through a~bifurcating tube,
resembling the problems studied in hemodynamics.
In~this last example, the artificial boundaries are 
planes placed such that their normal vector coincides with the tube axis (and with the expected flow).
In this case, the inflow is imposed by Dirichlet boundary conditions, while the
{\em do-nothing} condition is prescribed on two ``outlet'' endings.
Analogously to the previous examples, 
we find two distinct steady solutions, depending on the initial guess assigned to the Newton's method;
one of the solutions displays rapid inflow through the smaller ending.

\section{Isotropic radial flow}\label{sec:radial}
The multiple solutions described in this paper follow the mechanism
which is well illustrated by the following example.
The velocity field of the steady isotropic radial planar flow can be written in the form 
\newcommand{\nocolsep}{{\hskip-\arraycolsep}}
\newcommand{\vectxy}{
	\vect{x}
}
$$	\vect{u} = U(r) \vectxy
,\qquad	\textrm{with } r=|\vectxy| %
,\;\;	\vectxy=(x,y)
.$$
The constraint of incompressibility then implies that 
\begin{equation}\label{rad:div}
	0 = \divv \vect{u} = 2U + rU'
\qquad	\implies 
\qquad	U(r) = r^{-2}Q 
,\end{equation}
for a~constant $Q\in\mathbb{R}$.
Here and in what follows we do not indicate that $U$, $U'$ are functions of 
$r$.
The convective term takes the form
\begin{equation}\label{rad:conv}
	\divv ( \vect{u}\otimes\vect{u} ) 
=	( 3U^2 + 2rUU' ) \vectxy
\overset{\eqref{rad:div}}{=}
	- r^{-4} \, Q^2 \vectxy
.\end{equation}
Moreover, since
$$	\nabla \vect{u} 
=	(\nabla\vect{u})^T
=	\frac1r \left(\begin{array}{cc}	rU +  x^2 U'	&  xyU'	\\
					xy U'		& rU +  y^2 U'
		\end{array}\right)	
\overset{\eqref{rad:div}}{=}
	\frac{Q}{r^4} \left(\begin{array}{cc}	r^2 - 2x^2 	&  -2xy	\\
						-2xy 		& r^2 - 2y^2
		\end{array}\right)	
,$$
there holds
\begin{equation}\label{rad:divS}
	\divv( \nu\nabla\vect{u} )
=	\divv( \nu\,(\nabla\vect{u})^T )
=	\nu\,( 3 r^{-1} U' + U'' ) \vectxy
\overset{\eqref{rad:div}}{=}
	0
.\end{equation}
Thus, the balance of momentum in \eqref{eq:NS} %
allows us to write the pressure field as
$$	p = P(r),
\qquad	\textrm{so that}
\qquad	\nabla p = r^{-1} P' \vectxy
,$$
and is reduced to the following equation for the two unknowns $U\equiv U(r)$ and $P\equiv P(r)$,
\begin{equation}\label{rad:momentum}
	0 
= 	( 3U^2+2rUU' ) - 2\nu\,(3r^{-1}U' +U'') + r^{-1}P' 
\overset{\eqref{rad:div}}{=}
	-r^{-4}\,Q^2 + r^{-1}P' 
,\end{equation}
implying that 
\begin{equation}\label{rad:p}
	P' = r^{-3} \,Q^2
\qquad	\implies
\qquad	P = P_\infty - \frac12 r^{-2} Q^2
,\quad	\textrm{with some } P_\infty\in\mathbb{R}
.\end{equation}
Note that the viscous term does not appear here and that the pressure is 
an~increasing function of the radius for any nonzero flux $Q$, 
no matter of its sign, i.e.\ for both the inwards or outwards direction of the flow.

Let us define the domain $\Omega$ and its boundary 
$\partial\Omega = \Gamma_1\cup\Gamma_2\cup\Gamma_{\mathrm{n}}$
as follows, for given radii $0<R_1<R_2$ and given $0<\alpha\leq\pi/2$,
\begin{equation}\label{omega}
\left.\begin{aligned}
	\Omega	&=	\{ R_1^2 < x^2+y^2 < R_2^2 \,,\; x>0 \,,\; 0 < y/x < \tan(\alpha) \}
,\\	\Gamma_i &=	\partial\Omega \cap \{ x^2+y^2 = R_i^2 \} \,,\; i=1,2
,\\	\Gamma_{\mathrm{n}} &=	\partial\Omega \cap ( \{ x=y/\tan(\alpha) \} \cup \{ y=0 \} )
.\end{aligned}\right.
\end{equation} %
On $\Gamma_{\mathrm{n}}$, the outer normal $\vect{n}$ is orthogonal to the vector $(x,y)$,
implying that $(\nabla\vect{u})\vect{n} = (\nabla\vect{u})^T\vect{n} = U\,\vect{n}$,
so that $-\tens{T}\vect{n} = (P-2\nu U)\,\vect{n}$.
The radial flow \eqref{rad:div}--\eqref{rad:momentum} thus meets 
the perfect-slip boundary conditions 
(denoting by $\vect{w}_\tau = \vect{w}-(\vect{w}\cdot\vect{n})\vect{n}$ the tangential part of a~vector)
\begin{equation}\label{rad:slip}
	\vect{u}\cdot\vect{n} = 0 
\quad	\textrm{and}
\quad	(-\tens{T}\vect{n})_{\tau} = \vect{0}
\qquad	\textrm{on } \Gamma_{\mathrm{n}}
.\end{equation}
On $\Gamma_1$ and $\Gamma_2$, the outer normal $\vect{n} = (\pm1/r)\vectxy$
is parallel to the direction of the flow.
Therefore,
$	(\nabla\vect{u})\vect{n} 
=	(\nabla\vect{u})^T\vect{n} 
=	( U + r U' )\vect{n}
$ and the observed traction is given by
\begin{equation}\label{rad:traction-gen}
	(-\tens{T}\vect{n})\cdot\vect{n}
= 	P - 2\nu\,(U +rU')
\quad	\textrm{and}
\quad	(-\tens{T}\vect{n})_\tau = \vect{0}
\qquad	\textrm{on } \Gamma_1 \cup \Gamma_2
.\end{equation}
In particular, for the incompressible fluid, \eqref{rad:div} and \eqref{rad:p}
lead to the following relation for the normal traction, 
\begin{equation}
\label{rad:traction}
	-\tens{T}\vect{n}\cdot\vect{n}
=	P_\infty - \frac12 \frac{Q^2}{r^2} + 2\nu \frac{Q}{r^2} 
=	P_\infty - \frac{Q}{r^2} \left( \frac{Q}2 - 2\nu \right) 
\quad	\textrm{on }\Gamma_1\cup\Gamma_2
,\end{equation}
where $r=R_i$ on $\Gamma_i$, $i=1,2$.

\subsection{Nontrivial steady solution for trivial data}
As a~consequence of \eqref{rad:traction}, 
the incompressible Navier--Stokes equations in $\Omega$
subject to the perfect-slip boundary conditions \eqref{rad:slip} and 
the zero {\em traction} boundary conditions \eqref{eq:b}
$$	-\tens{T}\vect{n} = \vect{b} \equiv \vect{0}
\qquad	\textrm{on } \Gamma_1\cup\Gamma_2
$$
exhibit two isotropic radial solutions:
one trivial, defined by $P_\infty=Q=0$, 
and one given by 
$$	P_\infty = 0
\qquad	\textrm{and}
\qquad	Q = 4\nu
.$$
In the case that the zero {\em traction} conditions are replaced by the {\em do-nothing} conditions \eqref{eq:bhat}
\begin{equation*}
	p \vect{n} - \nu\,(\nabla\vect{u})\vect{n} = \hat{\vect{b}} \equiv 0
\qquad	\textrm{on } \Gamma_1\cup\Gamma_2
,\end{equation*}
since by simple computation
$	p\vect{n} - \nu\,(\nabla\vect{u})\vect{n}
=	- \tens{T}\vect{n} + \nu\,(\nabla\vect{u})^T\vect{n}
=	( P_\infty - \tfrac12 r^{-2} Q^2 + \nu r^{-2} Q )\,\vect{n}
$,
one makes the similar observation,
the non-trivial solution being defined by 
$$	P_\infty = 0
\qquad	\textrm{and}
\qquad	Q = 2\nu
.$$
The resulting flux observed for the do-nothing conditions is half of the flux
for the zero {\em traction} conditions, cf.~\eqref{eq:estimate}.

\subsection{Two, one or no solution for given {$P_i$--drop}}
\label{sec:twooneorno}
Let us further consider the {$P_i$--drop problem}%
\footnote{%
By the phrase {$P_i$--drop} we avoid using the more common, but in our setting confusing, phrase {\em pressure drop}.
	See the footnote in \Section{sec:structure}.
}, where the flow is subject to 
two values of the normal traction prescribed on $\Gamma_1$ and $\Gamma_2$,
\begin{equation}\label{rad:pressuredrop}
	-\tens{T}\vect{n} 
=	\vect{b}
\equiv 	P_i \,\vect{n}
\qquad	\textrm{on } \Gamma_i,
\quad	i = 1,2
.\end{equation}
For the given radii $R_i$ and the normal tractions $P_i$, 
one obtains by subtracting \eqref{rad:traction} for $i=1,2$,
the following quadratic equation for the unknown $Q$,
$$	P_1-P_2
=	-\, (R_1^{-2}-R_2^{-2}) (\tfrac12 Q^2 - 2\nu\,Q )
.$$
Having $Q$ in hand, $P_\infty$ is computed by using \eqref{rad:traction} for either $i=1$ or $2$. %

For $P_1-P_2 < \Pcrit$, there are two solutions for $Q$ as follows,
\begin{equation}\label{eq:twoQs}
	Q = 2\nu \left( 1 \pm \sqrt{ 
		1 - \frac{ P_1-P_2 }{ \Pcrit } 
		} \right)
,\qquad
	\Pcrit = 2 \nu^2\, (R_1^{-2}-R_2^{-2})
.\end{equation}
It is worth noting that %
the larger of the two solutions $Q$ is always positive and that it is a~{\em decreasing} function of 
the difference $P_1-P_2$.
For the special case $P_1-P_2 = \Pcrit$,
there is only one solution, given by 
$Q = 2\nu$.
Interestingly, there is no steady isotropic radial solution once the {$P_i$--drop} 
is too large, $P_1-P_2 > \Pcrit$.

In the case of {\em full-gradient-traction} conditions 
\begin{equation}\label{rad:bhatdrop}
	p\vect{n} - \nu\,(\nabla\vect{u})\vect{n}
=	\hat{\vect{b}}
\equiv 	P_i \,\vect{n}
\qquad	\textrm{on } \Gamma_i,
\quad	i = 1,2
,\end{equation}
one observes the analogous result, the critical {$P_i$--drop} being then at
\begin{equation}\label{eq:hatPcrit}
	\hatPcrit = \frac{\nu^2}{2}(R_1^{-2}-R_2^{-2})
,\end{equation}
and the unique solution being then given by $Q=\nu$, for $P_1-P_2 = \hatPcrit$.

\subsection{Stability of the one-dimensional problem}\label{sec:stability}
Let us now consider the unsteady flow of the same type,
i.e.\ let us find  
$$	\vect{u} = U(t,r) \vectxy
\qquad	\textrm{and}
\qquad	p = P(t,r)
$$
that solve the Navier--Stokes system of equations 
\newcommand{\xx}{} %
\begin{equation}\label{eq:unsteady}
\left.\begin{array}{rcl}
	\divv_{\xx}\vect{u} &=& 0 
\\	%
	\partial_t\vect{u}
	+ \divv_{\xx}(\vect{u}\otimes\vect{u}) %
-	\divv_{\xx}( \nu\,( \nabla\vect{u} + (\nabla\vect{u})^T) )
+	\nabla_{\xx} p
&=&	\vect{0} 
\end{array}\right\}
\textrm{ in $(0,T)\times\Omega$}
\end{equation}
supplemented by the (constant in time) boundary conditions \eqref{rad:slip}
and \eqref{rad:pressuredrop} or \eqref{rad:bhatdrop}.
Same as in the stationary case, the incompressibility constraint 
implies 
$$	U(t,r) = r^{-2}Q(t)
$$	
and, by using \eqref{rad:conv} and \eqref{rad:divS} the momentum equation implies that
$$	\partial_{r} P(t,r)
=	- r^{-1} Q'(t) + r^{-3} Q(t)^2 
,$$
which gives the pressure field 
$$	P(t,r) = P_\infty(t) - \ln(r)\,Q'(t) - \tfrac12 r^{-2} Q(t)^2  
.$$

Applying the constant $P_i$--drop by imposing \eqref{rad:pressuredrop},
one obtains from \eqref{rad:traction-gen}
the following two equations for the two unknown functions $P_\infty(t)$ and $Q(t)$, 
$$	P_i 
=	P_\infty - \ln(R_i) Q' - \tfrac12 R_i^{-2} Q^2  + 2\nu\, R_i^{-2} Q
,\qquad	i = 1,2
.$$
By subtracting, the following single equation for $Q(t)$ is concluded,
\begin{equation}\label{rad:ODE}
	a Q' = ( Q - b )^2 - c
,\quad	\textrm{where}
\quad	\left\{\begin{array}{rl}
	a \nocolsep&= \frac{ 2 \ln(R_2/R_1)}{ R_1^{-2}-R_2^{-2}} > 0
,\\	b \nocolsep&= 2\nu > 0
,\\	c \nocolsep&%
		=	4\nu^2\left( 1 - \frac{P_1-P_2}{\Pcrit} \right)
,\end{array}\right.
\end{equation}
with $\Pcrit$ from \eqref{eq:twoQs}.
Its solution $Q(t)$ appears in three different forms, depending on the sign of $c$.

For $c>0$, i.e.~for $P_1-P_2<\Pcrit$, the two stationary solutions $Q=b\pm\sqrt{c}$ are indeed found
same as in the previous subsection, while all other solutions are of the form
$$	Q = b - \sqrt{c} \,\tanh\left( \frac{\sqrt{c}}{a}(t - t_0) \right)^{\pm1}
,$$
with $t_0\in\mathbb{R}$ to be determined by the initial condition.
In particular, we conclude the following asymptotics of the maximal solutions starting from $Q(0)$,
$$	\begin{array}{rll}
	\lim_{t\to+\infty} Q(t) \nocolsep& = b - \sqrt{c}
\qquad&	\textrm{for any }
	Q(0) < b + \sqrt{c}
,\\	\lim_{t\to t_0^-} Q(t) \nocolsep& = +\infty
\qquad&	\textrm{for any }
	Q(0) > b + \sqrt{c}
.\end{array}
$$
We have found that one of the stationary solutions is asymptotically stable within the considered class of radial solutions, 
while the second one is unstable.
In particular, in the case of zero $P_i$--drop,
the obvious trivial solution $Q\equiv0$
is asymptotically stable in the above sense, 
while the other solution given by $Q\equiv4\nu$ is unstable.

In the special case $c=0$, i.e. for $P_1-P_2=\Pcrit$,
there is the single stationary solution $Q\equiv2\nu$,
while the other solutions are 
$$	Q = b - \frac{a}{t-t_0}
,$$
with $t_0\in\mathbb{R}$.
In this case, 
$$	\begin{array}{rll}
	\lim_{t\to+\infty} Q(t) \nocolsep& = b
\qquad&	\textrm{for any }
	Q(0) < b 
,\\	\lim_{t\to t_0^-} Q(t) \nocolsep& = +\infty
\qquad&	\textrm{for any }
	Q(0) > b 
,\end{array}
$$
again with $0 < t_0 < +\infty$ depending on $Q(0)$.

Finally, for $c<0$ ($P_1-P_2>\Pcrit$), where no stationary isotropic radial solution can be found,
all solutions are in the form
$$	Q = b + \sqrt{-c} \, \tan\left( \frac{\sqrt{-c}}{a}(t - t_0) \right)
.$$
For any initial value, $Q(t)$ will blow up to $+\infty$ in finite time
(obviously less than $\tfrac{ \pi a }{ \sqrt{-c} }$).

Same as before, in the case of the {\em full-gradient-traction} $P_i$--drop \eqref{rad:bhatdrop}, 
the analogous observations hold with the only difference that 
$b$ and $c$ in~\eqref{rad:ODE} are to be replaced by 
$$	b = \nu
\qquad	\textrm{and} \qquad
	c = \nu^2 - \frac{ 2 (P_1-P_2) }{ R_1^{-2}-R_2^{-2} }
	  = \nu^2 \left( 1 - \frac{P_1-P_2}{\hatPcrit} \right)
,$$
with $\hatPcrit$ from \eqref{eq:hatPcrit}.
We note in particular the case of {\em do-nothing} conditions on both boundaries, i.e.\ $P_1=P_2=0$,
where the flow decays to zero if $Q(0)<2\nu$, while it blows up in finite time for any 
initial $Q(0)>2\nu$.

\section{Numerical simulations}\label{sec:fem}
We performed a~number of numerical simulations of the flows subject to 
the constant {\em traction} or {\em full-gradient-traction} conditions 
on opposite artificial boundaries. %
In what follows we present some of them, on some we report only briefly,
so that the presentation is confined within a~reasonable extent.
The numerical results were obtained by the finite element scheme
implemented using FEniCS \cite{AlnaesBlechta2015a,LoggMardalEtAl2012a}. %
\onlyME{\FIXME{codes?}}
Navier--Stokes equations were discretized using Taylor--Hood finite element pair
on a~triangular mesh within the mixed formulation,
i.e.\ we solve for $(\vect{u}_h,p_h) \in V_h\times Q_h$ such that
(taking $\vect{f}=\vect{0}$, for simplicity)
\begin{equation}\label{steady-traction}
	( \divv(\vect{u}_h\!\otimes\!\vect{u}_h) , \vect{w} )_\Omega
+	( \tilde{\tens{S}}(\nabla\vect{u}_h) , \nabla\vect{w} )_\Omega
-	( p_h , \divv\vect{w} )_\Omega
-	( q , \divv\vect{u}_h )_\Omega
=	- (\tilde{\vect{b}} , \vect{w} )_{\partial\Omega}
\end{equation}
holds for all $(\vect{w},q) \in V_h\times Q_h$,
cf.~\eqref{eq:NSb} and \eqref{eq:NSbhat}.
Either the constant traction \eqref{rad:pressuredrop} 
is prescribed, $\tilde{\vect{b}} \defeq \vect{b}=P_i\vect{n}$ on $\Gamma_i$, in which case 
$\tilde{\tens{S}}(\nabla\vect{u}_h) \defeq \nu\,(\nabla\vect{u}_h + (\nabla\vect{u}_h)^T)$,
or the weak formulation with the full gradient is used, 
$\tilde{\tens{S}}(\nabla\vect{u}_h) \defeq \nu \nabla\vect{u}_h$,
and the corresponding conditions \eqref{rad:bhatdrop} are imposed, 
prescribing again $\tilde{\vect{b}} \defeq \hat{\vect{b}}=P_i\vect{n}$ on $\Gamma_i$.

The Taylor--Hood finite element function spaces 
$V_h \subset W^{1,2}(\Omega)^2$ and $Q_h \subset L^2(\Omega)$
are defined in a~standard way, see e.g.~\cite{LoggMardalEtAl2012a}.
On the remaining parts of the boundary we prescribe either
(in Sections \ref{sec:fem-radial} and \ref{sec:fem-unsteady})
the perfect-slip boundary conditions \eqref{rad:slip}, %
in which case 
$V_h \subset \{ \vect{v}\in W^{1,2}(\Omega)^2 \,,\; \tr(\vect{v}\cdot\vect{n}) = 0 \textrm{ on } \Gamma_{\mathrm{n}} \}$;
or (in Sections \ref{sec:fem-JH}--\ref{sec:bif}) 
the no-slip condition on $\Gamma_0\subset\partial\Omega$, taking 
$V_h \subset \{ \vect{v}\in W^{1,2}(\Omega)^2 \,,\; \tr\,\vect{v} = \vect{0} \textrm{ on } \Gamma_{0} \}$.

While focusing on steady problems, we present a~few unsteady simulations as well.
The unsteady problems (in Sections \ref{sec:fem-unsteady} and \ref{sec:JH-unsteady}) are 
first discretized by the fully implicit (backward Euler) scheme with the fixed time step $\tau>0$,
i.e.\ we solve subsequently at each time level $t_k=k\tau$, $k=1,2,\ldots$, 
the following equation for the unknowns 
$(\vect{u}_h^k,p_h^k) \equiv (\vect{u}_h(t_k),p_h(t_k)) \in V_h \times Q_h$,
\begin{multline}\label{unsteady-traction}
	\frac1\tau \, ( \vect{u}_h^k , \vect{w} )_\Omega
+	( \divv(\vect{u}_h^k\otimes\vect{u}_h^k) , \vect{w} )_\Omega
+	( \tilde{\tens{S}}(\nabla\vect{u}_h^k) , \nabla\vect{w} )_\Omega
-	( p_h^k , \divv\vect{w} )_\Omega
-	( q , \divv\vect{u}_h^k )_\Omega
\\=	- ( \tilde{\vect{b}}, \vect{w} )_{\partial\Omega}
+ 	\frac1\tau \, ( \vect{u}_h^{k-1} , \vect{w} )_\Omega 
,\end{multline}
for all $(\vect{w},q) \in V_h\times Q_h$,
where $\vect{u}_h^0 = \vect{u}_h(0)$ is a~given initial condition.
Both the steady and the unsteady problem result in (one, or a sequence of) nonlinear algebraic systems,
that are solved by the FEniCS built-in DOLFIN Newton solver \cite{LoggMardalEtAl2012a,LoggWells2010a} 
with the exact Jacobian matrix computed by automated differentiation.
All numerical results presented were checked to be independent of a~further refinement in $h$ or $\tau$.

\subsection{Steady isotropic radial setting}\label{sec:fem-radial}
\figureSIRSabcd
We start by performing numerical simulations in the setting corresponding to the steady isotropic radial flow
discussed in \Section{sec:twooneorno}.
We define the flow domain~$\Omega$ as in~\eqref{omega} with $\alpha=\pi/2$,
so that the constraint $\vect{u}\cdot\vect{n}=0$ on $\Gamma_{\mathrm{n}}$ 
can be written, denoting $\vect{u}=(u,v)$, as  
$$	v = 0 \textrm{ on } \Gamma_{\mathrm{n}}\cap\{y = 0\}
\qquad	\textrm{and}
\qquad	u = 0 \textrm{ on } \Gamma_{\mathrm{n}}\cap\{x = 0\}
,$$
which simplifies the implementation. We set $R_1=1$, $R_2=3$ and $\nu=1$.

Prescribing the constant {\em traction} \eqref{rad:pressuredrop}, $\vect{b}=P_i\vect{n}$ on $\Gamma_i$,
for a~set of $P_i$--drops 
$P_1-P_2$, we report 
the resulting normal flux%
\footnote{
	Note that $\vect{e}_r = \frac1r(x,y) = -\vect{n}$ on $\Gamma_1$ while $\vect{e}_r = \vect{n}$ on $\Gamma_2$, 
	and that the definition of $Q_{\mathrm{num}}$ is independent on whether $i=1$ or $2$, due to $\divv\vect{u}=0$.
	For the isotropic radial flow presented in \Section{sec:radial}, where $\vect{u}=Q\tfrac1r\vect{e}_r$,
	there is $Q_{\mathrm{num}}=Q$.
}
$$	Q_{\mathrm{num}} = \frac1\alpha \int_{\Gamma_i} \vect{u}\cdot\vect{e}_r \,\mathrm{d}S
$$
in \Figure{fig:steady-traction-Qs}.
We make the following observation.
\begin{itemize}
\item	For $-1<P_1-P_2\leq1.\overline{7}$,
	by starting Newton's method from the zero initial guess
	we obtain the isotropic radial solutions,
	recovering the lower branch of \eqref{eq:twoQs}.
	Note that the critical {$P_i$--drop} with only one corresponding radial solution
	derived in \Section{sec:twooneorno} is 
	$\Pcrit= 2\nu^2\,(R_1^{-2}-R_2^{-2})=16/9=1.\overline{7}$.
\item	Starting Newton's method from the velocity field with large enough radial flux,
	specifically starting it from the isotropic radial solution $\vect{u}=r^{-2} Q_{\mathrm{init}} \vect{x}$ 
	with $Q_{\mathrm{init}}=3$,
	for the same range of {$P_i$--drops} as above %
	we obtain %
	different numerical solutions,
	recovering the upper branch of \eqref{eq:twoQs}.
\item	For $P_1-P_2$ larger than but close to the critical value $1.\overline{7}$,
	Newton's iterations (started by, for instance, $Q_{\mathrm{init}}=0$ or $Q_{\mathrm{init}}=3$)
	get lost. 
	The algebraic residuum oscillates and does not show much hope for convergence. %
\item	For the {$P_i$--drop} given large enough, say $P_1-P_2\geq2$, 
	in which case there exists no isotropic radial solution, 
	we are finding non-radial numerical solutions. %
	For convenience, we present a~simple ``non-radiality indicator''
	$$	i_{\mathrm{nr}} = \frac{ \| \vect{u}\cdot\vect{e}_{\theta} \|_{2,\Omega} }{ \| \vect{u} \|_{2,\Omega} }
	,\qquad	\textrm{where}\quad 
		\vect{e}_{\theta} = \tfrac1r (-y,x) 
	$$
	in \Figure{fig:steady-traction-rads}.
	By naive numerical continuation, namely by decreasing the {$P_i$--drop} subsequently
	while using the previous non-radial solution as the initial guess,
	we are finding the non-radial solutions for $P_1-P_2$ down to $1.65$, see \Figure{fig:steady-traction-Qs}.
	Note on \Figure{fig:steady-traction-rads} that the non-radiality gradually vanishes with decreasing {$P_i$--drop},
	the solutions eventually joining the lower branch of isotropic radial solutions.
\end{itemize}
\figureSIRSdonot

Numerical simulations using the {\em full-gradient-traction} \eqref{rad:bhatdrop}, 
$\hat{\vect{b}}=P_i\vect{n}$ on $\Gamma_i$,
lead to analogous results, as reported in \Figure{fig:steady-bhat-Qs} and \ref{fig:steady-bhat-rads}.
\begin{itemize}
\item
	Note that the critical {$P_i$--drop} here equals $\Pcrit=\tfrac12 \nu^2\,(R_1^{-2}-R_2^{-1})=4/9=0.\overline{4}$.
\item
The main difference regards the non-radial solutions found 
(e.g., first from the zero initial guess for $P_1-P_2=2$, and then numerically continued to lower $P_i$--drops).
As shown in \Figure{fig:steady-bhat-Qs}, the non-radial solutions now live down to the negative $P_1-P_2<-1.3$.
\item
In particular, we thus found three%
\footnote{%
	Or four, actually: the non-radial solutions appear in two variants symmetric around the axis $x=y$.
} %
distinct numerical solutions for trivial data, i.e. for the {\em do-nothing} boundary condition 
prescribed on~$\Gamma_0$ and $\Gamma_1$.
The velocity and pressure fields of the corresponding non-radial solution 
are shown in \Figure{fig:bhat-nonrad}.
\end{itemize}

\subsection{Unsteady isotropic radial setting}\label{sec:fem-unsteady}
Similarly as in Section~\ref{sec:stability}, we are interested in the unsteady flows 
in the very same setting as above, i.e.\ we keep the geometry, the boundary conditions and the viscosity 
same as in the previous section, but we look for the unsteady solutions to \eqref{eq:unsteady}
with the given initial state
\begin{equation*}\label{eq:initial-bc}
	\vect{u}(t=0) = \vect{u}_0
\quad	\textrm{in $\Omega$.}
\end{equation*}
We only perform simulations with stationary boundary conditions.

First, we set the initial velocity field simply as the isotropic radial velocity 
\begin{equation}\label{eq:initial-Qo}
	\vect{u}_0 = r^{-2} Q_0 \vect{x}
.\end{equation}
Running the numerical simulations for the $P_i$--drops $P_1-P_2 = -1$, $0$,\ldots,$3$
and $Q_0 = -1$, $0$,\ldots,$5$, we always reproduced the exact unsteady isotropic radial 
solutions derived in Section~\ref{sec:stability}.
Note that we stop the simulations at $t=10$, 
or if the velocity is becoming large, namely as soon as $Q_{\mathrm{num}}\geq20$.

In particular, the numerical solutions remain isotropic radial and show the 
blow-up in finite time for $P_1-P_2>\Pcrit$; 
while for $P_1-P_2<\Pcrit$, where there exist two steady solutions, the numerical solutions either 
converge asymptotically to the smaller (stable) steady isotropic radial solution, or they blow
up in finite time, depending %
on the initial state.
\onlyME{neradialni vynechat.}

\figureJHab
\figureSJHdonot
\figureHemo

\subsection{Steady {$P_i$--drop} diverging channel flow} \label{sec:fem-JH}
Inspired by the behavior of isotropic radial flows,
we now focus on the simulations which are not restricted to such simplified setting. 
We start, for simplicity, by keeping the domain geometry $\Omega$ similar as before,
only with $\alpha=\pi/8$ in~\eqref{omega}, while we keep $R_1=1$ and $R_2=3$.
We now prescribe the no-slip boundary conditions on the straight walls,
$$	\vect{u}=\vect{0} \quad\textrm{ on }\Gamma_0\equiv\Gamma_{\mathrm{n}}
.$$
We set the viscosity $\nu=1/20$, so that the resulting fluxes of the two branches of 
solutions will be of the same order as in \Section{sec:fem-radial}.
We report them in \Figure{fig:steady-JH}, %
showing the behavior analogous to the one presented in \Figure{fig:steady-traction-bhat}. %
The results for {\em traction} and {\em full-gradient-traction} boundary data are now plotted together,
showing that 
\begin{itemize}
\item	again, the initial guess based on $Q_{\mathrm{init}}=3$ reveals 
	the upper branch of steady solutions;
\item	the resulting flux on the upper branch for given {\em traction} 
	is larger than the one for {\em full-gradient-traction}, 
	although the difference is less pronounced than it was for the isotropic radial flow;
\item	no numerical steady solutions are obtained for large enough $P_i$--drops.
\end{itemize}

For illustration, we present in \Figure{fig:bhat-jh} the velocity and pressure fields of the 
non-trivial steady numerical solution for zero {\em full-gradient-traction} $P_i$--drop,
i.e.\ with the {\em do-nothing} boundary conditions.
It shows that the pressure is nearly zero in the vicinity of the artificial boundaries, as expected,
while it is lower inside the domain. 
The pressure field indicates that the velocity cannot be radial.
The flux is around $Q_{\mathrm{num}}=2.7$, in accordance with \Figure{fig:steady-JH}.

\subsection{Unsteady $P_i$--drop diverging channel solutions}\label{sec:JH-unsteady}
Analogously to \Section{sec:fem-unsteady}, we look for the unsteady solutions,
while keeping the geometry, boundary conditions and viscosity the same as in \Section{sec:fem-JH}.
For simplicity, we set the initial velocity field numerically as the isotropic 
radial velocity \eqref{eq:initial-Qo} (despite the fact that it does not satisfy the 
no-slip boundary conditions).
Running the numerical simulations for $Q_0=-1$, $0$, \dots, $4$
and various $P_i$--drops, we obtain results analogous to those in \Section{sec:fem-unsteady}.
We report in \Figure{fig:UnsteadyJH} the solutions for three {\em full-gradient-traction} $P_i$--drops.
For $P_1-P_2=0$, all but one solutions converge to zero, 
while the one starting from $Q_0=4$ displays the blow-up in finite time. 
Similarly for $P_1-P_2=20$, the solutions either converge to the lower steady solution,
or blow up if the initial-state flux is larger than that of the upper steady solution.
For $P_1-P_2=30$, all the solutions blow up, the fluxes $Q_{\mathrm{num}}(t)$ appear to 
be only shifted in time. 

\subsection{Flow through a~bifurcating tube}\label{sec:bif}
Finally, we study the three-dimensional flow in the~geometry of a~bifurcating tube,
let us say a bifurcating blood vessel,
see \Figure{fig:hemo}.
The inflow through one inlet boundary (on the top) is given,
by prescribing the parabolic velocity profile via Dirichlet boundary condition,
the no-slip conditions are assumed on the walls.
The outflow is possible through two outlets (on the bottom), 
where we set the {\em do-nothing} boundary condition.
The geometry of the domain was constructed such that the artificial boundaries represent circular cross-sections 
of the vessel parts, the diameters of the inlet and the two outlets being $0.5$, $0.2$ and $1.0$.
The length of the domain (from the inlet to the outlets) is $3$ and the distance of the outlets centers is approximately~$1$, 
the inflow velocity is $1$ at its midflow maximum and the viscosity%
\footnote{The quantities we work with can be considered as non-dimensional here, 
	the viscosity thus represents a~reciprocal of Reynolds number. 
	Note that 
	Reynolds number $\mathrm{Re}=50$ is related to the characteristic length equal to the diameter of the larger outlet,
	or to the distance of the centers of the two outlets, and to the characteristic velocity equal to the peak velocity at the inlet.
	Note that the peak backflow velocity of the second solution reaches five times more, 
	which would then correspond to $\mathrm{Re}=250$.
} is given by $1/\nu = 50$.

The fact that the cross-sectional area of the two outlets differ significantly is important for the existence of 
two steady laminar numerical solutions.
Figures~\ref{fig:hemo-a} and \ref{fig:hemo-b} present two steady flows subject to the same
boundary conditions: the no-slip condition on the walls and the {\em do-nothing} on the outlets.
The non-trivial flux through the inlet was prescribed only in order to make the example more representative
for the real application; 
prescribing the trivial data on the inlet as well would result in the very same non-unique behaviour.

\subsection{Comment on other examples}\label{sec:other}
Analogously, one can obtain more than two solutions in domains with more than two artificial 
boundaries treated with the conditions \eqref{eq:b} or \eqref{eq:bhat}, provided that the ratios
of their area are large enough and the domain allows for flows with small enough dissipation.
One such example, which we do not report in detail here, would consist of a triangular domain, 
where \eqref{eq:b} or \eqref{eq:bhat} is given on certain parts (of a~different length) 
of each edge of the triangle.
For certain lengths of the artificial boundaries,
it is easy to find combinations of $P_i$--data allowing for up to three (reasonably small, laminar)
steady solutions.
Such a~setting, however, is not of much practical importance.

On the other hand, by for example gradually prolonging the outlets in \Figure{fig:hemo} by straight tubes,
while watching the resulting inflow through the smaller outlet of the solution illustrated in \Figure{fig:hemo-b}, 
one would observe a~significant increase of the inflow velocity, 
loosing soon the possibility of finding two (laminar) steady solutions.
This is in accordance with the fact, that the {\em do-nothing} conditions are used in practical
simulations on regular basis without reporting multiple solutions such as those considered in this paper.
According to our experience, it is only in specific applications that one can encounter this non-uniqueness accidentally.

\section{Conclusion}
The fact that the boundary conditions prescribing a~constant value of {\em traction} or
what we call the {\em full-gradient-traction} here, %
including the {\em do-nothing} boundary conditions, inhibit from establishing the well-posedness theory
for the steady flow problem,  
is well known by the mathematical community concerned with the %
flows of incompressible fluids.
The banality with which the {\em do-nothing} conditions are used 
in a~multitude of numerical simulations none the less, 
reflects %
the amenity of these boundary conditions and their robustness in many practical flow problems. 
It is not the aim of this paper whatsoever to reason against using them, 
rather we attempt to contribute to the discussion on how to use them and what results can be expected. 

Clearly, following the examples presented in this paper, the {\em uniqueness} of the steady solution to the problems subject to boundary conditions
that include \eqref{eq:b} or \eqref{eq:bhat} cannot be expected in general, not even for small or trivial data. 
Although this may not be surprising, it has not been reported in the literature in this detail before, 
to the best of our knowledge.
The more prominent question, whether multiple solutions can be found in problems 
where {\em only one} connected part of the boundary is treated in this way, remains unresolved. 
We can only report without any details that we have tried to find such numerical examples as well, 
yet with no success. 

Inspired by Sections~\ref{sec:fem-unsteady} and \ref{sec:JH-unsteady}, other questions arise as well:
in the case of multiple steady solutions (say, for small data), is there still a~unique 
{\em stable} steady solution, as both the numerical simulations in the diverging channel 
and the analytical isotropic radial solutions suggest?
In particular, is the {\em unique small} solution established by Galdi in~\cite{Galdi2008-inHemodynamicalFlows}
the one (and only one) that is stable?
Somewhat related is the question, which of the multiple solutions can be delivered by various
non-linear algebraic problem solvers, as we only reported the numerical experiments with Newton's method
(which can reveal multiple solutions, depending on the initial guess), 
and the solutions to the unsteady flow (which, if considered as a~tool for finding the steady solution,
either failed or revealed one steady solution at most).

Other presented examples suggest the lack of {\em existence} of the steady solution 
for certain problems with large enough boundary data. 
Such behavior of Navier--Stokes equations %
is expected in general, but largely due 
to the onset of an unstable non-laminar flow regime,
in which case the existence of the numerical solution can be affected 
by various convective term stabilization techniques. 
Here, the isotropic radial example reveals a~rather different mechanism 
related to the blow-up of the unsteady flow in finite time due to the kinetic energy 
being brought into the domain by the inflow through one of the artificial boundaries.
It seems obvious that this behavior can only be avoided by altering the boundary conditions. 

\bibliographystyle{unsrtnat}
\footnotesize
\bibliography{selected}

\end{document}